# Trends of epitaxial perovskite oxide films catalyzing the oxygen evolution reaction in alkaline media

*Denis Antipin,[1] Marcel Risch[1,2,*]*

*[1]Nachwuchsgruppe Gestaltung des Sauerstoffentwicklungsmechanismus, Helmholtz-Zentrum Berlin für Materialien und Energie GmbH, Hahn-Meitner-Platz 1, 14109 Berlin*
*marcel.risch@helmholtz-berlin.de*

*[2]Georg-August Universität Göttingen, Institut für Materialphysik, Friedrich-Hund-Platz 1, 37077 Göttingen*

**Abstract**

The oxygen evolution reaction (OER) is considered a key reaction for electrochemical energy conversion but slow kinetics hamper application in electrolyzers, metal-air batteries and other applications that rely on sustainable protons from water oxidation. In this review, the prospect of epitaxial perovskite oxides for the OER at room temperature in alkaline media is reviewed with respect to fundamental insight into systematic trends of the activity. First, we thoroughly define the perovskite structure and its parameter space. Then, the synthesis methods used to make electrocatalytic epitaxial perovskite oxide are surveyed and we classify the different kinds of electrodes that can be assembled for electrocatalytic investigations. We briefly discuss the semiconductor physics of epitaxial perovskite electrodes and their consequences for the interpretation of catalytic results. OER investigations on epitaxial perovskite oxides are comprehensively surveyed and selected trends are graphically highlighted. The review concludes with a short perspective on opportunities for future electrocatalytic research on epitaxial perovskite oxide systems.

## 1. Introduction

Water splitting is the most important reaction for sustainable production of hydrogen equivalents (protons) [1–4]. These are used in one half reaction to make energy carriers (e.g., $H_2$, $CH_3OH$) [5–11], fertilizers ($NH_3$) [12–14] or feedstocks for the chemical industry (e.g., $CH_2O$) [15–17]. The supply of the much needed hydrogen equivalents is limited by the half reaction of oxygen evolution from water, also known as water oxidation [18–20]. The oxygen evolution reaction (OER) suffers from large energy losses due to a high overpotential, which is rooted in the mechanism of transferring four electrons and four hydroxide molecules (in alkaline media) [21–24]. While much of our current knowledge of likely mechanisms derives from theoretical work [21,23,25,26], very few details of the mechanism are supported experimentally. A key concept in the rational design of better electrocatalysts is the so-called descriptor approach, i.e., property activity relationships [23,27,36,28–35]. The latter calls for systematic investigations of defined materials with tunable properties such as epitaxial perovskite oxides.

The family of perovskite oxides is a popular choice for systematic investigations because the composition of the perovskites can be varied with little change to the structural framework in the best case [32,37–40]. In the last decade, the interest in epitaxially deposited perovskite oxides has increased considerably [41–44]. The preparation of epitaxial thin films and buried solid–solid interfaces since the late 1980s has led to a very mature understanding of the solid-state chemistry and oxide physics of perovskite oxides [45,46]. These films have vast potential to further advance our understanding of the solid-liquid interface due to the controlled orientation of the films, their negligible roughness and the absence of additives such as carbon or a binder that are commonly added to composite electrodes based on powders. Therefore, epitaxial perovskite oxides are the perfect choice for systematic investigations of electrocatalytic reactions and in particular of the ill-understood OER.

The scope of this review is a comprehensive overview of OER studies on epitaxial perovskite oxides. Firstly, the perovskite structure is rigorously and quantitatively defined. Then, synthesis methods for these films and the resulting physical properties are discussed and we survey how electrodes we made from the as-deposited films. The next section deals with charge transfer across buried solid-solid interfaces and across the solid-liquid interface, which must not limit catalysis. Subsequently, trends in the published works are identified and discussed based on the fundamentals introduced in earlier sections. In particular, we analyze how well an electronic and a structural descriptor describe the available activities of epitaxial perovskite oxides. The review is concluded by a summary of the current state of the small but burgeoning field and a perspective for future research directions.

## 2. What is a perovskite?

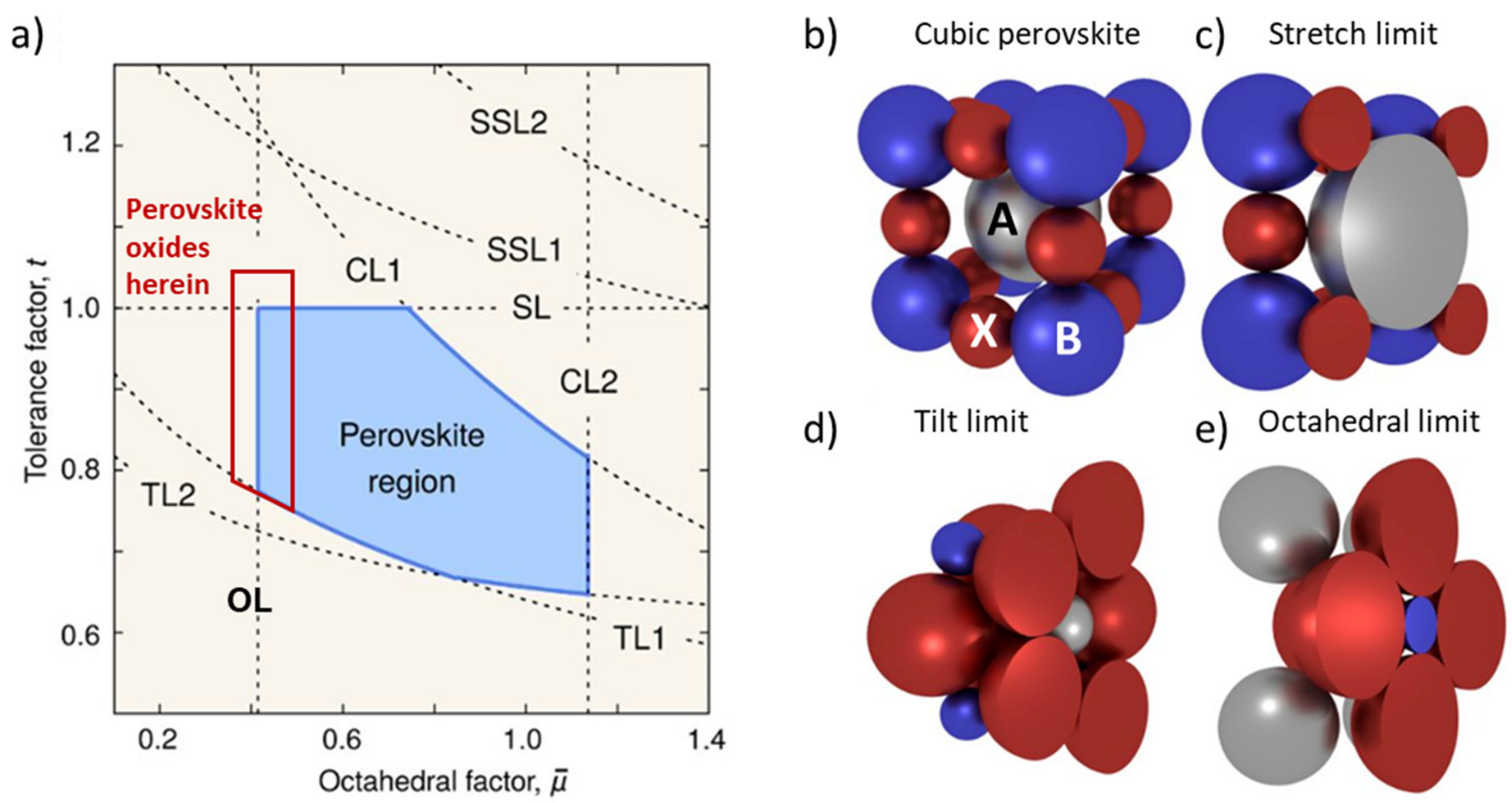

**Fig. 1.** (a) Theoretically derived stability region (solid blue area) of general perovskites $ABX_3$ in the parameter space defined by the tolerance and octahedral factors. The red outlined area indicates the parameter space of the electrocatalytic perovskite oxides studied herein. The shown limits are: chemical limits (CL1 and CL2), octahedral limit (OL), stretch limits (SL, SSL1 and SSL2), and tilt limits (TL1 and TL2). Further detail may be found in the text. Accompanying illustrations of (b) a cubic perovskite, (c) the stretch limit due to large A-site, (d) the tilt limit due to small A-sites, and (e) the octahedral limit due to the relative size of the B-site and oxygen. Modified with permission from ref. [47]. Copyright 2018 National Academy of Sciences.

Perovskite oxides denotes a family of compounds with general structure $ABO_3$ named after the mineral $CaTiO_3$ which defines the prototypical structure [48]. Commonly, lanthanides or group II elements are found on the A-site and transition metals on the B-site. The term “perovskite” is currently used ambiguously even though a precise definition exists. According to Breternitz and Schorr [49], three structural requirements must be met that define a general perovskite (including non-oxides):

R1) A stoichiometry of $ABX_3$, or at least a ratio A:B:X of 1:1:3
R2) The coordination of the B-site cation needs to be octahedral (or a distorted octahedra)
R3) The $[BX_6]$ octahedra need to be organized in an all-corner sharing 3D network

The authors justify these rules based on the relationship between structure and properties where the properties would differ strongly, especially if the network of corner-sharing octahedra was lost. The ratio in requirement R1 includes double perovskites with general formula $A2BB'X_6$, for which this network is preserved. Herein, we would like to rephrase point 1 to "a ratio of A:B:X of near 1:1:3" to include materials with anionic or cationic vacancies that still fulfill requirements R2 and R3, e.g., oxygen-vacancy ordered double perovskites with formula $AA'B_2O_{5+\delta}$ or regular perovskites (fulfilling R2 and R3) that have sizable number of vacancies, which often are important for the electronic and structural properties.

The fulfilment of the above requirements naturally leads to quantitative definitions and requirements based on ionic radii. Goldschmidt [50] defined the tolerance factor as

$$t = \frac{r_A + r_O}{\sqrt{2}(r_B + r_O)} \qquad (1)$$

where $r_A$, $r_B$ and $r_O$ are the ionic radii of ions at the A-site, B-site and oxygen, respectively. Most commonly, specific values are taken from Shannon's Table [51] or linearly extrapolated from these tabulated values to the appropriate coordination.

Li, Soh and Wu [52] found that the Goldschmidt tolerance factor is not sufficient to predict the formability of the perovskite structure and they introduced the octahedral factor as a secondary requirement. The octahedral factor (of perovskite oxides) is defined as:

$$\mu = \frac{r_B}{r_O} \qquad (2)$$

These definitions and their significance can be rationalized geometrically. The A-site cation exactly fits in the void between the B-O octahedra if the B-site cations sit on the corners of a cube and the size of the A-site cation matches the diagonal. This leads to a cubic bulk perovskite with tolerance factor very close to unity ($t \approx 1$; Fig. 1b). The octahedral cavity forms a regular cuboctahedron and the A-site is 12-fold coordinated. If the A-site cation is larger, it does not longer fit in the octahedral void. This scenario is denoted SL (stretch limits) in Fig. 1a and illustrated in Fig. 1c. In this case, requirements R2 and R3 needs to be broken and non-perovskite crystal structures are observed. If the A-site is smaller than the octahedral cavity, the B-O octahedra tilt, rotate and often distort, which strongly affects the hybridization of electronic orbitals. Moreover, the A-site is moved out of the center of the octahedral void. The coordination of the A-site is reduced in this configuration. This scenario is denoted TL (tilt limits) in Fig. 1a and illustrated in Fig. 1d. In these limits, the oxygen ions of neighboring octahedra touch and again requirement 3 cannot be not fulfilled. The octahedral factor is limited by the case when the anions touch, i.e. when the B-site radius is small compared to the ionic radius of oxygen. It is denoted OL (octahedral limit) in Fig. 1a and illustrated in Fig. 1c. This limit is observed if $r_B = r_O\left(1 - \frac{1}{\sqrt{2}}\right) \leq 0.41$ Å, which is smaller than the ionic radii of 6-fold coordinated first row transition metals [51]. However, stable perovskites were also found to the left of this boundary [47]. Finally, the upper boundary of the perovskite stability region in Fig. 1a is given by the ratio of the ionic radii of Cs (A-site) and F as the anion as well as $Fr^+$ and $Ac^{3+}$ as B-sites, which is denoted CL (chemical limit).

For the perovskite oxides reviewed here as electrocatalysts for the OER, the largest A-site cation is Ba, the anion is O and the largest B-site is $Ir^{3+}$. The latter gives an upper octahedral factor of $\mu = 0.486$ and thus the stretch limit (SL) rather than chemistry (i.e. CL1) limits the tolerance factor to $t$ = 1.0 for electrocatalytic perovskite oxides. The calculated tolerance factors of the perovskite oxides discussed herein range from $t$ = 0.958 to $t$ = 1.053 (Table 1). About a third (28 %) of all entries have a tolerance factor above $t_{SL}$ = 1.0. The actual tolerance factors based on experimental bond distances may differ slightly. The limits discussed above are usually not strict and stable perovskite oxides that fulfil the requirements R1-3 can be observed experimentally also slightly outside the predicted ranges. The calculated octahedral factors range from $\mu$ = 0.371 to $\mu$ = 0.456, where again the lower boundary is very clearly outside the predicted stability boundary of $\mu_{OL} > (\sqrt{2}-1) \approx 0.41$ [47]. In fact, 77% of all entries in Table 1 have an octahedral factor below $\mu_{OL}$ and 28 % of all entries have both octahedral factors below $\mu_{OL}$ and tolerance factors above $t_{SL}$, meaning all entries with $\mu < \mu_{OL}$ also show $t > t_{SL}$. Filip and Giustino [47] attribute their outliers to the simplicity of the used Goldschmidt model and found that values outside their predicted stability ranges may indicate polymorphism.

In summary, the criteria for definition of perovskites are well defined. Yet, many perovskite oxides studied as catalysts for the OER fall outside the stability range predicted by the Goldschmidt model where a third (28 %) of all investigated perovskites are found in the top left corner of the red area and thus outside of the Goldschmidt model. While these make up only a small percentage of all possibly stable perovskites, a better model would be desirable for those perovskite oxides used as electrocatalysts for the OER. Additionally, tolerance and octahedral factors based on measured bond distances would be likewise desirable, particularly for epitaxial films where strain between the substrate and film may lead to locally different values.

**Table 1.** Comprehensive overview of structural properties, fabrication methods and activity of 53 epitaxial films used as catalysts for the OER and additional films where the activity could not be evaluated.

| Material | Tolerance factor (*t*) | Octahedral factor (*μ*) | Substrate/ Support | Synthesis method | Electrode type* | Conductivity (S/cm) | Orientation | E at 50 $\mu A/cm^2_{ox}$ (V vs RHE) | j at 1.6 V vs RHE ($\mu A/cm^2_{ox}$) | Tafel slope (mV/dec) | Electrochemical method | Reference |
|---|---|---|---|---|---|---|---|---|---|---|---|---|
| $SrIrO_3$ | 0.992 | 0.446 | $DyScO_3$ | MBE | Rectangular masked | n/a | (100) | 1.53 | 910 | 41 | CA, 0 rpm | [53] |
| $NdNiO_3$ | 0.963 | 0.400 | $LaSrAlO_4$ | PLD | Rectangular masked | $1.3*10^{4}$ [a] | (100) | 1.61 | 39 | 68 | CA, 0 rpm | [54] |
| $NdNiO_3$ | 0.963 | 0.400 | $LaAlO_3$ | PLD | | $2.2*10^{4}$ [a] | (100) | 1.65 | 16 | 100 | CA, 0 rpm | [54] |
| $NdNiO_3$ | 0.963 | 0.400 | $NdGaO_3$ | PLD | | $1.5*10^{4}$ [a] | (100) | 1.66 | 7 | 132 | CV, 0 rpm | [54] |
| $NdNiO_3$ | 0.963 | 0.400 | $SrTiO_3$ | PLD | | $4.2*10^{3}$ [a] | (100) | 1.64 | 9 | 139 | CV, 0 rpm | [54] |
| $LaCoO_3$ | 0.996[b] | 0.400 | $SrTiO_3$ | PLD | Rectangular masked | $1.2*10^{3}$ [c] | (100) | 1.58 | 114 | 55 | CA, 0 rpm | [55] |
| $LaCoO_3$ | 0.996[b] | 0.400 | $(LaAlO_3)_{0.3}(Sr_2AlTaO_6)_{0.7}$ | PLD | | $1.2*10^{3}$ [c] | (100) | 1.58 | 148 | 64 | CA, 0 rpm | [55] |
| $LaCoO_3$ | 0.996[b] | 0.400 | $LaAlO_3$ | PLD | | $4.25*10^{2}$ [c] | (100) | 1.60 | 48 | 109 | CA, 0 rpm | [55] |
| $LaNiO_3$ | 0.996 | 0.400 | $LaAlO_3$ | OPA-MBE | Rectangular masked | $8*10^{3}$ [a] | (100) | 1.60 | 19 | 141 | CV, 0 rpm | [56] |
| $La_{0.88}Sr_{0.12}NiO_3$ | 1.004 | 0.393 | $LaAlO_3$ | OPA-MBE | | $4.3*10^{3}$ [a] | (100) | 1.63 | 28 | 136 | CA, 0 rpm | [56] |
| $La_{0.75}Sr_{0.25}NiO_3$ | 1.013 | 0.386 | $LaAlO_3$ | OPA-MBE | | $2.7*10^{3}$ [a] | (100) | 1.60 | 47 | 112 | CA, 0 rpm | [56] |
| $La_{0.5}Sr_{0.5}NiO_3$ | 1.031 | 0.371 | $LaAlO_3$ | OPA-MBE | | $2*10^{3}$ [a] | (100) | 1.56 | 151 | 90 | CA, 0 rpm | [56] |
| $LaCoO_3$ | 0.996[b] | 0.400 | $SrTiO_3$ | PLD | Rectangular masked | 0.8 [a] | (100) | ---[d] | 16[d] | 154 | CV, 0 rpm | [57] |
| $La_{0.8}Sr_{0.2}CoO_3$ | 1.005[b] | 0.396 | $SrTiO_3$ | PLD | | $5.2*10^{2}$ [a] | (100) | 1.58 | 94 | 84 | CV, 0 rpm | [57] |
| $La_{0.6}Sr_{0.4}CoO_3$ | 1.013[b] | 0.391 | $SrTiO_3$ | PLD | | $2.6*10^{3}$ [a] | (100) | 1.53 | 360 | 81 | CV, 0 rpm | [57] |
| $La_{0.5}Sr_{0.5}CoO_3$ | 1.018[b] | 0.389 | $SrTiO_3$ | PLD | | $1.8*10^{3}$ [a] | (100) | 1.55 | 200 | 78 | CV, 0 rpm | [57] |
| $La_{0.4}Sr_{0.6}CoO_3$ | 1.022[b] | 0.387 | $SrTiO_3$ | PLD | | $3.4*10^{2}$ [a] | (100) | 1.57 | 74 | 161 | CV, 0 rpm | [57] |
| $LaNiO_3$ | 0.996 | 0.400 | $Nb:SrTiO_3$ | PLD | Rectangular masked | $2.3*10^{4}$ [e] | (100) | 1.64 | 14 | 81 | CA, 0 rpm | [58] |
| $La_{0.5}Nd_{0.5}NiO_3$ | 0.979 | 0.400 | $Nb:SrTiO_3$ | PLD | | $1.3*10^{4}$ [e] | (100) | 1.63 | 24 | 109 | CA, 0 rpm | [58] |
| $La_{0.2}Nd_{0.8}NiO_3$ | 0.970 | 0.400 | $Nb:SrTiO_3$ | PLD | | $3.4*10^{3}$ [e] | (100) | 1.61 | 40 | 103 | CA, 0 rpm | [58] |
| $NdNiO_3$ | 0.963 | 0.400 | $Nb:SrTiO_3$ | PLD | | $2.5*10^{3}$ [e] | (100) | 1.61 | 33 | 67 | CA, 0 rpm | [58] |
| $Nd_{0.5}Sm_{0.5}NiO_3$ | 0.958 | 0.400 | $Nb:SrTiO_3$ | PLD | | $1.7*10^{3}$ [e] | (100) | 1.60 | 41 | 74 | CA, 0 rpm | [58] |
| $La_{0.6}Sr_{0.4}MnO_3$ | 0.988 | 0.428 | $Nb:SrTiO_3$ | IBS | Disk assembly | $2.4*10^{3}$ [a] | (100) | 1.64 | 6 | 63 | CV, 2500 rpm | [59] |
| $La_{0.6}Ca_{0.4}CoO_3$ | 0.999[b] | 0.391 | MgO | PRCLA | Not reported | $1.5*10^{2}$ [a,g] | (100) | 1.58[f,h] | 53[f,h] | n/a | CV, 0 rpm | [60] |
| $La_{0.6}Ca_{0.4}CoO_3$ | 0.999[b] | 0.391 | MgO | PRCLA | | $1.5*10^{2}$ [a,g] | (100)+(110) | 1.68[f,h] | 22[f,h] | n/a | CV, 0 rpm | [60] |
| $La_{0.6}Ca_{0.4}CoO_3$ | 0.999[b] | 0.391 | MgO | PRCLA | | $1.5*10^{2}$ [a,g] | (110) | 1.65[f,h] | 40[f,h] | n/a | CV, 0 rpm | [60] |
| $LaNiO_3$ | 0.996 | 0.400 | $LaSrAlO_4$ | PLE | Disk masked | $5.4*10^{3}$ [a] | (100) | 1.63 | 19 | 56 | CV, 1600 rpm | [61] |
| $LaNiO_3$ | 0.996 | 0.400 | $LaAlO_3$ | PLE | | $6.0*10^{3}$ [a] | (100) | 1.62 | 21 | 46 | CV, 1600 rpm | [61] |
| $LaNiO_3$ | 0.996 | 0.400 | $(LaAlO_3)_{0.3}(SrAl_{0.5}Ta_{0.5}O_3)_{0.7}$ | PLE | | $2.4*10^{3}$ [a] | (100) | 1.63 | 15 | 68 | CV, 1600 rpm | [61] |
| $LaNiO_3$ | 0.996 | 0.400 | $SrTiO_3$ | PLE | | $1.6*10^{3}$ [a] | (100) | 1.66 | 8 | 65 | CV, 1600 rpm | [61] |
| $LaNiO_3$ | 0.996 | 0.400 | $DyScO_3$ | PLE | | $1.1*10^{3}$ [a] | (100) | 1.69 | 2 | 72 | CV, 1600 rpm | [61] |
| $Ba_{0.5}Sr_{0.5}Co_{0.8}Fe_{0.2}O_{3-\delta}$ on $La_{0.8}Sr_{0.2}MnO_3$ (0%)[i] | 1.053[b,j] | 0.404 | $Nb:SrTiO_3$ | PLD | Rectangular masked | n/a | (100) | 1.68 | 4 | 69 | CV, 0 rpm | [62] |
| $Ba_{0.5}Sr_{0.5}Co_{0.8}Fe_{0.2}O_{3-\delta}$ on $La_{0.8}Sr_{0.2}MnO_3$ (48%)[i] | 1.053[b,j] | 0.404 | $Nb:SrTiO_3$ | PLD | | n/a | (100) | 1.54 | 920 | 50 | CV, 0 rpm | [62] |
| $Ba_{0.5}Sr_{0.5}Co_{0.8}Fe_{0.2}O_{3-\delta}$ on $La_{0.8}Sr_{0.2}MnO_3$ (94%)[i] | 1.053[b,j] | 0.404 | Nb:$SrTiO_3$ | PLD | | n/a | (100) | 1.51 | 2500 | 53 | CV, 0 rpm | [62] |
| $Pr_{0.5}Ba_{0.5}CoO_{3-\delta}$ | 1.044[b] | 0.400 | $SrTiO_3$ | PLD | Rectangular masked | $3.6*10^{3}$ [e] | (100) | 1.69[f] | ---[d] | n/a | CV, 0 rpm | [63] |
| $(Pr_{0.5}Ba_{0.5})_2Co_2O_{5.5+\delta}$[m] | 1.044[b] | 0.400 | $SrTiO_3$ | PLD | | $7.4*10^{2}$ [e] | (100) | 1.67[f] | ---[d] | n/a | CV, 0 rpm | [63] |
| $La_{0.6}Sr_{0.4}CoO_3$ | 1.013[b] | 0.391 | $NdGaO_3$ | PLD | Rectangular masked | $3.3*10^{3}$ [n] | (100) | 1.48[f] | 1280 | n/a | CV, 0 rpm | [64] |
| $La_{0.88}Sr_{0.12}FeO_3$ | 0.961 | 0.456 | Nb:$SrTiO_3$ | OPA-MBE | Rectangular masked | $3.3*10^{-2}$ [n] | (100) | 1.58 | 97 | n/a | CV, 0 rpm | [65] |
| $SrRuO_3$ | 0.994 | 0.443 | Nb:$SrTiO_3$ | RFMS | Disk assembly | n/a | (100) | 1.43[o] | ---[d] | n/a | CV, 1600 rpm | [66] |
| $SrRuO_3$ | 0.994 | 0.443 | Nb:$SrTiO_3$ | RFMS | | n/a | (110) | 1.32[o] | ---[d] | n/a | CV, 1600 rpm | [66] |
| $SrRuO_3$ | 0.994 | 0.443 | Nb:$SrTiO_3$ | RFMS | | n/a | (111) | 1.30[o] | ---[d] | n/a | CV, 1600 rpm | [66] |
| $Pr_{0.9}Ca_{0.1}MnO_3$ | 0.964 | 0.453 | Nb:$SrTiO_3$ | IBS | Disk assembly | n/a | (100) | 1.67 | 4 | n/a | CV, 0 rpm | [67] |
| $Pr_{0.67}Ca_{0.33}MnO_3$ | 0.974 | 0.434 | Nb:$SrTiO_3$ | IBS | | n/a | (100) | 1.64 | 14 | n/a | CV, 0 rpm | [67] |
| $LaNiO_3$ | 0.996 | 0.400 | $SrTiO_3$ | Sol-gel | Rectangular masked | n/a | (100) | 1.61 | 55 | 69 | CV, 0 rpm | [68] |
| $LaNiO_3$[p] | 0.996 | 0.400 | $SrTiO_3$ | Sol-gel | | n/a | (100) | 1.56 | 575 | 40 | CV, 0 rpm | [68] |
| $PrNiO_3$ | 1.001 | 0.400 | $SrTiO_3$ | Sol-gel | | n/a | (100) | 1.59 | 60 | 72 | CV, 0 rpm | [68] |
| $PrNiO_3$[p] | 1.001 | 0.400 | $SrTiO_3$ | Sol-gel | | n/a | (100) | 1.56 | 580 | 37 | CV, 0 rpm | [68] |
| $NdNiO_3$ | 0.963 | 0.400 | $SrTiO_3$ | Sol-gel | | n/a | (100) | 1.59 | 55 | 72 | CV, 0 rpm | [68] |
| $NdNiO_3$[p] | 0.963 | 0.400 | $SrTiO_3$ | Sol-gel | | n/a | (100) | 1.55 | 792 | 36 | CV, 0 rpm | [68] |
| $La_{0.6}Sr_{0.4}MnO_3$ | 0.988 | 0.428 | Nb:$SrTiO_3$ | IBS | Disk assembly | n/a | (100) | 1.64 | 13 | n/a | CV, 1600 rpm | [69] |
| $LaMnO_3$ | 0.954 | 0.461 | Nb:$SrTiO_3$ | PLD | Not reported | n/a | (100) | 1.72 | ---[d] | n/a | CV, 0 rpm | [28] |
| $SrRuO_3$[q] | 0.994 | 0.443 | Nb:$SrTiO_3$ | PLE | Rectangular masked | $4.5*10^{-3}$ [n] | (100) | 1.36 | n/a | 78 | CV, 0 rpm | [70] |
| $SrRuO_3$[r] | 0.994 | 0.443 | Nb:$SrTiO_3$ | PLE | Rectangular masked | $3.6*10^{-3}$ [n] | (100) | 1.23 | n/a | 109 | CV, 0 rpm | [70] |
| $LaCoO_3$ | 0.996[b] | 0.400 | $LaAlO_3$ | Sol-gel | Disk masked | $1.5*10^{2}$ [a] | (100) | --- [d,f] | --- [d,f] | ---[d] | CV, 0 rpm | [71] |
| $LaCoO_3$ | 0.996[b] | 0.400 | $LaAlO_3$ | Sol-gel | | $2.3*10^{2}$ [a] | (110) | --- [d,f] | --- [d,f] | ---[d] | CV, 0 rpm | [71] |
| $LaCoO_3$ | 0.996[b] | 0.400 | $LaAlO_3$ | Sol-gel | | $3.6*10^{2}$ [a] | (111) | --- [d,f] | --- [d,f] | ---[d] | CV, 0 rpm | [71] |
| $La_{0.8}Sr_{0.2}CoO_3$ | 1.005[b] | 0.396 | $SrTiO_3$ | PLD | Not reported | --- | (100) | --- [d,f] | --- [d,f] | ---[d] | CV, 0 rpm | [72] |
| $La_{0.8}Sr_{0.2}CoO_3$ | 1.005[b] | 0.396 | $SrTiO_3$ | PLD | | --- | (110) | --- [d,f] | --- [d,f] | ---[d] | CV, 0 rpm | [72] |
| $La_{0.8}Sr_{0.2}CoO_3$ | 1.005[b] | 0.396 | $SrTiO_3$ | PLD | | --- | (111) | --- [d,f] | --- [d,f] | ---[d] | CV, 0 rpm | [72] |
| $La_{0.8}Sr_{0.2}CoO_3$ | 1.005[b] | 0.396 | --- | FZ | Single crystal rod | --- | (100) | --- [d,f] | --- [d,f] | ---[d] | CV, 1600 rpm | [73] |
| $La_{0.8}Sr_{0.2}CoO_3$ | 1.005[b] | 0.396 | --- | FZ | | --- | (110) | --- [d,f] | --- [d,f] | ---[d] | CV, 1600 rpm | [73] |
| $La_{0.8}Sr_{0.2}CoO_3$ | 1.005[b] | 0.396 | --- | FZ | | --- | (111) | --- [d,f] | --- [d,f] | ---[d] | CV, 1600 rpm | [73] |
| $SrCoO_{3-\delta}$[s] | 1.036 | 0.384 | $LaSrAlO_4$ | PLE | Disk masked | 0.1 [a] | (100) | --- [d] | --- [d] | ---[d] | CV, 1600 rpm | [74] |
| $SrCoO_{3-\delta}$[s] | 1.036 | 0.384 | $(LaAlO_3)_{0.3}$ $(SrAl_{0.5}Ta_{0.5}O_3)_{0.7}$ | PLE | | 7.7 [a] | (100) | --- [d] | --- [d] | ---[d] | CV, 1600 rpm | [74] |
| $SrCoO_{3-\delta}$[s] | 1.036 | 0.384 | $SrTiO_3$ | PLE | | $1.1*10^{2}$ [a] | (100) | --- [d] | --- [d] | ---[d] | CV, 1600 rpm | [74] |
| $SrCoO_{3-\delta}$[s] | 1.034 | 0.387 | $DyScO_3$ | PLE | | $2.2*10^{2}$ [a] | (100) | --- [d] | --- [d] | ---[d] | CV, 1600 rpm | [74] |
| $SrCoO_{3-\delta}$[s] | 1.034 | 0.388 | $GdScO_3$ | PLE | | $6.0*10^{2}$ [a] | (100) | --- [d] | --- [d] | ---[d] | CV, 1600 rpm | [74] |
| $SrCoO_{3-\delta}$[s] | 1.033 | 0.389 | $KTaO_3$ | PLE | | $8.0*10^{2}$ [a] | (100) | --- [d] | --- [d] | ---[d] | CV, 1600 rpm | [74] |

[a] calculated from resistivity data at T = 298 K measured by PPMS; [b] calculated using $Co^{3+}$ intermediate state (IS) ionic radius [75]; [c] calculated from EIS data and the thickness of the film; [d] was not able to calculate or determined from presented data; [e] calculated from sheet resistance at T =298 K measured by Van der Pauw method; [f] 1M KOH solution; [g] a composition of the film is not clear from the original paper that was used for OER measurements, so the lowest value for resistivity was chosen for conductivity calculations; [h] recalculated to the RHE scale from Hg/HgO using indicated electrolyte concentration and $E_0$(Hg/HgO) = 118 mV; [i]

percentage of BSCF coverage is indicated in parentheses; [j] δ = 0.4 was used for calculations [27]; [k] the conductivity was determined by the measurement of potential drop between two probes attached to the (100) planes of the single crystal with the silver paste, during the d.c. constant current flow; [l] was normalized manually on the geometric surface area taken from the article (1 $mm^2$); [m] double-perovskite structure, formed by ordering of oxygen vacancies within every $CoO_2$ plane; [n] calculated from resistivity data at T = 298 K measured by Van der Pauw method; [o] overall current has an impact from Ru dissolution of ≈ 10%; [p] CVs were measured after surface Fe exchange; [q] orthorhombic structure; [r] tetragonal structure; [s] δ from the article was used for calculations [74].

Ionic radius for $Pr^{3+}$ with coordination number (CN) 12 was calculated by linearizing existing data for $Pr^{3+}$ (CNs 6, 8 and 9) and extrapolating needed radius value.

Abbreviations: RHE = reversible hydrogen electrode; CV = cyclic voltammetry; CA = chronoamperometry; rpm = revolutions per minute; MBE = molecular-beam epitaxy; OPA-MBE = oxygen-plasma-assisted MBE; PLD = pulsed-layer deposition; IBE = ion-beam sputtering; PRCLA = pulsed reactive crossed-beam laser ablation; PLE = pulsed layer epitaxy; LPE = liquid-phase epitaxy; FZ = floating-zone method; RFMS = radio-frequency magnetron sputtering, FZ = floating zone method.

## 3. Synthesis methods

Perovskite oxides can be epitaxially deposited by a multitude of thin film deposition methods. These can generally be classified by how the atoms for deposition are produced, namely using light, ballistics, heat or solvation, and by how these atoms react, namely, physically or chemically. The nature of production and deposition determines which substrates are suitable (e.g. conductive or insulating) and it affects the film properties and films of nominally identical composition may thus differ, e.g. in the type and concentration of defects or in morphology, when prepared by different methods. An in-depth discussion of epitaxial film deposition is beyond the scope of this review and the readers are referred to refs. [76–83].

The epitaxial oxides reported as catalysts for the OER were made by mainly pulsed laser deposition (PLD), pulsed laser epitaxy (PLE) or pulsed reactive crossed beam laser ablation (PRCLA), where the atoms are produced by laser ablation of a target and react physically with the substrate surface [81,84,85]. These methods amount to 64% of all entries in Table 1. The next most popular methods for producing thin films are based on sputtering, namely ion beam sputtering (IBS) and radio-frequency magnetron sputtering (RFMS). The atoms for deposition are produced ballistically by bombarding a target material with high energy ions in IBS. In RFMS, the ions of an argon plasma are accelerated by a radio-frequency potential towards the target and the sputtering yield is further increased by a magnetic field [76–79]. The atoms react physically upon deposition on the substrate. Epitaxial films were also made by molecular beam epitaxy (MBE) and oxygen plasma-assisted molecular beam epitaxy (OPA-MBE), where molecular beams are directed towards the substrate. The atoms for deposition are usually produced from thermally evaporated elemental target materials. An oxygen plasma can be used to prepare oxide films by physical reactions with the substrate [78,80].

Epitaxial thin films have also been made by chemical rather than physical methods. The sol-gel method can be used to obtain thin films with different orientations [68,71] The ions for deposition are dissolved homogeneously mixed soluble salts in solutions that are applied onto the substrate by spin-coating or similar techniques. The film forms upon chemical reaction of the solved ions with the substrate.

The wetting of the deposited ions on the substrate determines the microstructure and whether a closed film form. In a simple model, the surface energy may be such that it is minimized by deposition of ions on the used substrate or that it is minimized by grouping multiple deposited ions. The former case will give monolayers and layer growth if the spreading is still favored (Frank–van der Merwe growth) and the latter case will give island growth (Vollmer-Weber growth). Various microstructures have been reported for epitaxial oxides used as catalysts for the OER (Fig. 2). Most substrates are not cut exactly along the desired crystal orientation but have a slight misalignment. This produces a terraced surface, which usually requires an etching step for uniform sharp steps. In the best case, the epitaxially deposited film conforms to the well-defined morphology of the substrate (Fig. 2a). The growth of the epitaxial film may produce decorations on the terrace edges (Fig. 2b) or roughen the terrace edges (Fig. 2c). A smooth granular structure may also cover the (possibly present) terrace edges (Fig. 2d) or even produce a different microstructure, e.g. of micron-sized islands (Fig. 2e). The morphology is important for systematic investigations, where exposure of a single surface facet is highly desired. Any structural defect such as step edges, kinks, vacancies, adatoms or islands exposes secondary high-index facets. These defects, particularly step edges have also been proposed as active sites for metal electrocatalysts [86,87]. Despite their different morphologies, all films shown in Fig. 2 showed epitaxial growth by X-ray diffraction.

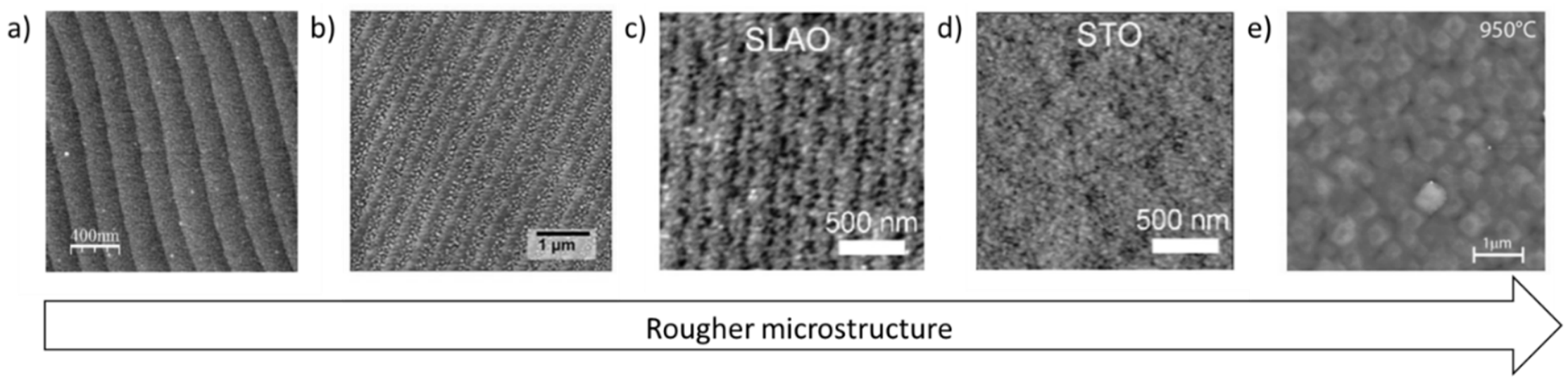

**Fig. 2.** Representative AFM images of epitaxial perovskite oxide surfaces showing layer growth with (a) sharp terraces due to miscut of the substrate, (b) decorated terraces, (c) rough granular terraces, (d) no terraces but a smooth granular surface, and (e) island growth. Note the different scale bars. Panels a, c, d, e were reproduced with permission from refs. [54,61,63]. Copyright American Chemical Society 2016, 2017, 2019. Panel b is reproduced under CC BY-NC 4.0 license from ref. [69].

The perovskite oxide films herein were deposited on mainly on $SrTiO_3$ (STO) with or without doping with Nb (often 0.5 wt%) to make the substrate conductive (66 % of all entries in Table 1). $SrTiO_3$ is a cubic perovskite ($t$ = 1.002 , $\mu$ = 0.432) with a lattice parameter of $a_{STO}$ = 3.905 Å [88]. Nb-doping changes this value insignificantly, e.g. ref. [89]. Many perovskite oxides with first row transition metal B-sites have similar pseudocubic lattice parameters and thus grow well on $SrTiO_3$, i.e. with small lattice mismatch. Pseudocubic means that the actual space group of the perovskite is orthorhombic or rhombohedral ($t$ < 1) but the unit cell is translated and rotated to resemble a cubic one [90]. This pseudocubic unit cell may further be rotated with respect to the cubic unit cell of $SrTiO_3$ to minimize lattice mismatch. Other substrates include $DyScO_3$ (DSO), $LaSrAlO_4$ (LSAO), $LaAlO_3$ (LAO), $NdGaO_3$ (NGO) and $(LaAlO_3)_{0.3}(SrAl_{0.5}Ta_{0.5}O_3)_{0.7}$ (LSAT), $(LaAlO_3)_{0.3}(Sr_2AlTaO_6)_{0.7}$ (LSAT), where the most common short hand notations are given in parentheses. These substrates are selected either to minimize the lattice mismatch (e.g. $La_{0.6}Sr_{0.4}CoO_3$ on $NdGaO_3$ [64]) or to study the effects of deliberate lattice mismatch, i.e. strain, where compressive strain increases the out-of-plane lattice parameter of the film and tensile strain decreases it.

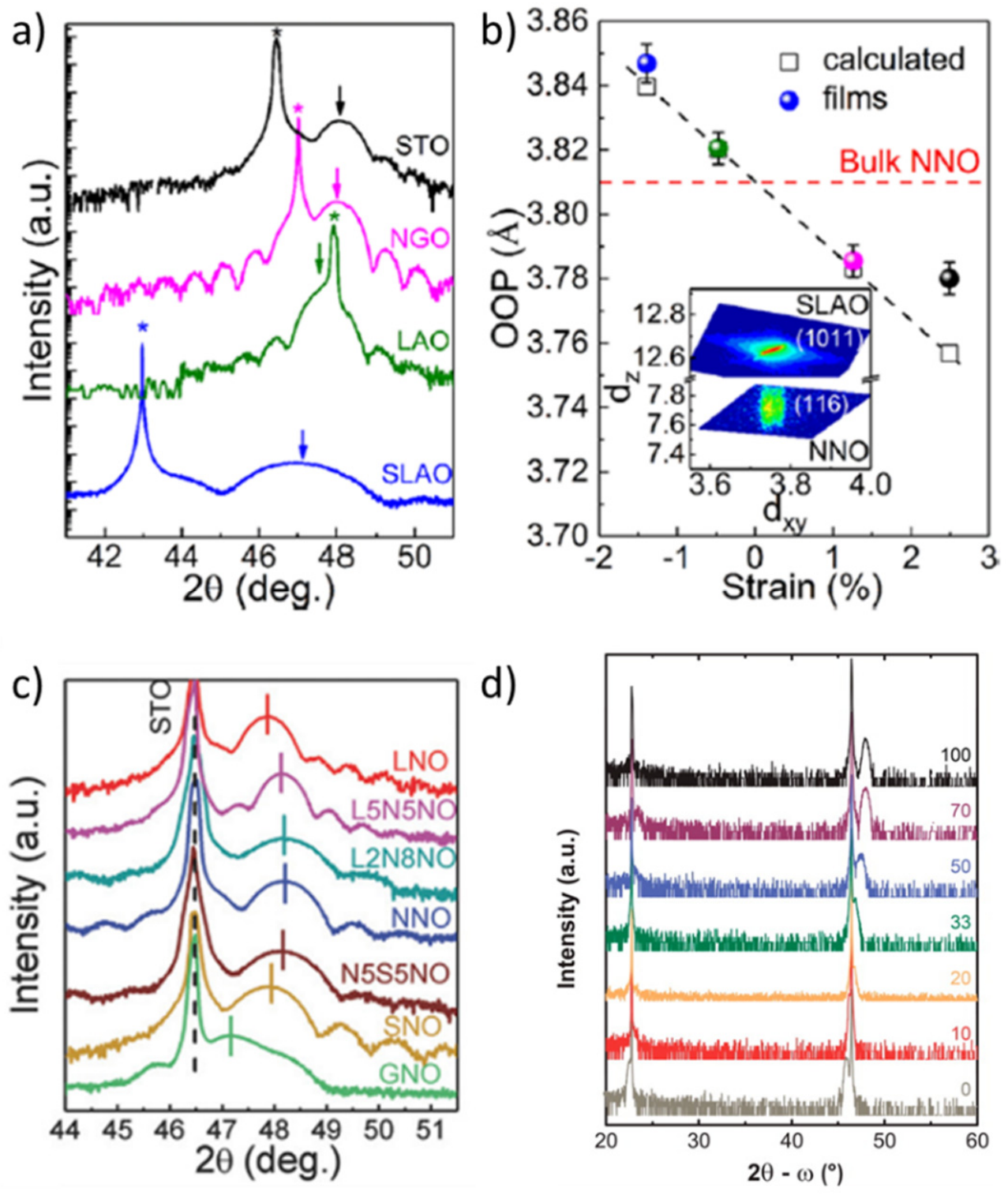

**Fig. 3.** (a) Out-of-plane (OOP) XRD scans of $NdNiO_3$ on $SrTiO_3$ (STO), $NdGaO_3$ (NGO), $LaAlO_3$ (LAO) and $SrLaAlO_4$ (SLAO). (b) Corresponding plot of the measured (solid symbols) and predicted (open squares) OOP lattice parameters as function of strain for these substrates. The inset shows a reciprocal space map. Additional OOP XRD scans of epitaxial oxides and their substrates for (c) the A-site cation of AA'$NiO_3$ on Nb:$SrTiO_3$ and (d) Sr doping of the A-site of $La_{1-x}Sr_xCoO_3$ on a buffer layer of $La_{0.8}Sr_{0.2}MnO_3$ on $SrTiO_3$. Panels a, b, d were reproduced with permission from refs. [54,91]. Copyright American Chemical Society 2015, 2019. Panel c republished with permission of John Wiley and Sons from ref. [58].

Exemplary out-of-plane θ–2θ scans for $NdNiO_3$ on various substrates are shown in Fig. 3a. The 002 reflection of $NdNiO_3$ is indicated by an arrow. A lower angle indicates a larger pseudocubic lattice parameter due to Bragg's law [92]. In Fig. 3b, the calculated experimental lattice parameters are compared to the bulk value of $NdNiO_3$ to determine if the strain is compressive (negative values) or tensile (positive values). The inset in Fig. 3b further shows a reciprocal space map of $NdNiO_3$ on $SrLaAlO_4$, which supports in-plane alignment of the (pseudocubic) lattice parameters of substrate and film. Strong strain effects can also be observed for selection of dissimilar sized A-site cations on (Fig. 3c). Yet, the extent of strain may be small for selection of similarly sized cations in the substrate and films as well as for their mixtures, e.g., La (1.36 Å) and Sr (1.44 Å ) [51], on a suitable substrate or buffer layer such $La_{0.8}Sr_{0.2}MnO_3$ for $La_{1-x}Sr_xCoO_3$ deposition on $SrTiO_3$ (Fig. 3d) [57]. Strain is an important parameter as it firstly determines whether epitaxy is possible at extreme values (depending on the composition of film and substrate) and at less extreme values what kind of distortions occur to reduce the surface energy of the mismatched interfaces. The coordination octahedra can distort and tilt (further), which affects bond lengths and angles. Electronically, the crystal field splitting is modified and the covalence and charge transfer between the B-site transition metal and oxygen may change. It has been shown that these properties affect the electrocatalysis of oxygen, which is discussed in more detail below [27,32,33,93–97]. Thus, epitaxially strained films offer a new degree of freedom for the design of electrocatalysts. The insight from these model systems may also be relevant, e.g., for core-shell nanoparticles or other bilayer or multilayer systems.

In summary, the film properties depend on the synthesis or deposition method where similar ion sources and deposition mechanisms are expected to produce similar properties. There is no systematic study yet how comparable nominally identical perovskite oxide films behave electrochemically, which would be a desirable multi-laboratory effort. For identical deposition methods, the growth mode depends on the composition and structural details of the substrate and deposited film. Various morphologies have been reported with different degrees of roughness that must match the desired research goal. Furthermore, epitaxial films offer another degree of freedom for the design of catalysts, namely the strain between the substrate and deposited film. The aspect has been well studied [54,55,61,74] and the influence on the OER has been attributed to resulting changes in crystal field splitting due to octahedral distortion [61] and bond angles, i.e. hybridization [54,74].

## 4. Assembly of electrodes for electrocatalytic investigations

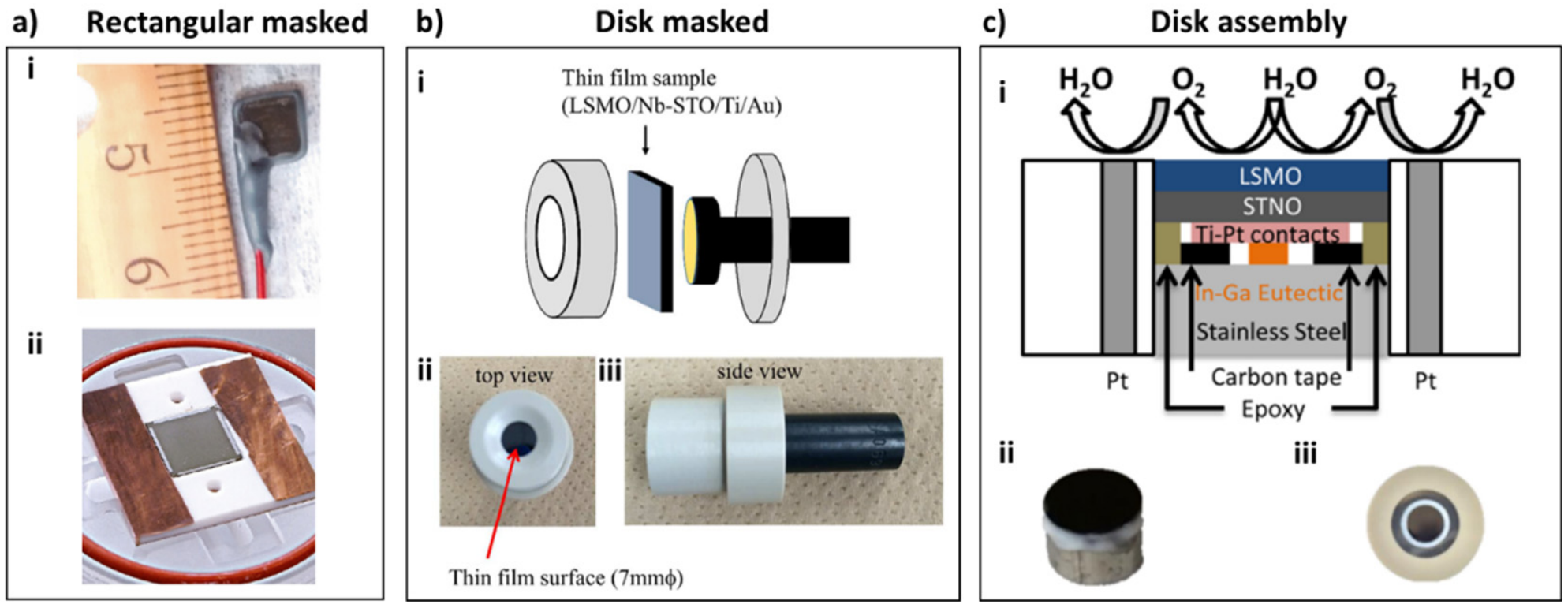

**Fig. 4.** Fabrication of electrodes for electrocatalytic experiments: (a) rectangular masked electrodes, (b) masked rotating disk electrodes and (c) rotating disk assemblies. Panel ai republished with permission of John Wiley and Sons from ref. [58]. Panels aii-c were reproduced with permission from refs. [59,64,98]. Copyright American Chemical Society 2016, 2019.

Different approaches have been taken to assemble electrodes from epitaxial thin films for electrocatalytic investigations. In order of increasing complexity of the assembly, the range from immersed rectangular electrode assemblies (Fig. 4a; “rectangular masked”) to square films in an assembly on a rotating disk (Fig. 4b; “disk masked”) to fabrication of rotating disks for commercial electrode holders (Fig. 4c; “disk assembly”). Most commonly, the substrate is connected to the working electrode lead of the potentiostat (“back contact”) and Ti/Au or Ti/Pt metal contacts are fabricated to ensure Ohmic contact with the electric contacts in the holders or with the used wire. Alternative connecting schemes are used when the substrate requires for epitaxy is not conductive. Details of the assembly and the advantages or disadvantages of the three approaches are discussed in the following paragraphs.

Rectangular masked electrodes are commonly made by attaching a suitable wire, often Ti, to an epitaxially deposited substrate (Fig. 4a i). This approach is used most frequently in electrocatalytic investigations with epitaxial oxides (70% of all entries in Table 1). In addition to the aforementioned back contact, the wire can also be connected to the active layer (“front contact”) or to both the active layer and the substrate (“side contact”). Initial fabrication steps often include cleaning of the contact area and wire and roughening of the contact areas. It should be noted that the latter procedure will likely expose the substrate and connect it electrically. Further fabrication steps include application of In-Ga eutectic to make an Ohmic contact between the oxide and metal, application of silver point for fixation and finally masking the substrate back, sides and contact area using non-conductive chemically resistant epoxy. In an alternative approach [64], the rectangular sample is not masked by epoxy but by clamping a cover onto the surface (Fig. 4a ii). In this example, a 6 mm O-ring (not shown) is used to define the electrochemical surface exposed to the electrolyte. The advantages of these electrodes are the ease of fabrication and the compatibility with standard rectangular substrates of sizes 5x5, 5x10 or 10x10 mm$^2$. The limitations are the lack of control over charge transport, which is much less of an issue for the OER as compared to other reactions such as the ORR, and the difficulty of performing spectroscopic or microscopic post-mortem characterizations on masked samples. Yet, the latter point is alleviated by clamping the sample.

Rectangular epitaxial films can also be adapted to a rotating disk either by gluing them to the surface (not shown) [61] or masking them with a circular aperture (Fig. 4b). We could not find further information on the glued samples but this approach likely did not yield a circular electrochemical area and the analytical expression of the limiting current may be unknown. The masked setups are used by the Kan and Chueh groups [98,99]. In the shown example, a 10x10 mm$^2$ substrate was masked by a cover with a 7 mm diameter aperture, which effectively converts the square film to a disk for electrocatalytic experiments. A back contact is the most natural connection in this approach. The home-made holder could then be mounted onto a conventional rotator and used akin to regular disks. While the OER in aqueous solutions is not limited by the reactants water or hydrogen, the forced convection by rotation removes bubbles from the surface. Additional advantages of this approach are the reusability of the epitaxial film for post-mortem investigations, and the control over mass transport. Disadvantages are possible contributions of stray resistance and capacitance of the covered film parts as well as the necessity to construct a holder.

Epitaxial films can also be deposited directly on circular substrates (Fig. 4c) [59,66,67,69]. This approach is used by the Markovich [66], Shao-Horn [59], Jooss [59,67,100] and Risch [69] groups . For the latter three groups, these epitaxially deposited substrate disks are then connected by In-Ga eutectic onto suitable conductive metal disks, e.g. made of stainless steel or aluminum. Additional fixation is provided by strips of double-sided carbon tape. The small gap between the support disk and the deposited substrate is finally filled with non-conductive epoxy. The disk electrodes assembled by this approach have been successfully mounted in commercial holders of rotating disk electrodes (RDE) and rotating ring-disk electrodes (RRDE) with 4 mm [59,67,69,100] and 5 mm diameter [59,66] in the same way a commercial metal disk or glassy carbon disk would be mounted. This approach also allows control over mass transport. Furthermore, the ring can be used to detect products such as oxygen, when mounted into an RRDE. The disadvantage is the complex assembly that only allows contacting on the back, which restricts it to conducting substrates, and the assembly complicates post-mortem analysis.

In summary, three different approaches for fabricating electrodes from epitaxial perovskite oxides have been discussed. Each has their own advantages and disadvantages as listed above, which can serve as a guide to new researchers in the field which preparation is best to achieve their research goals.

## 5. Semiconductor physics of epitaxial electrodes

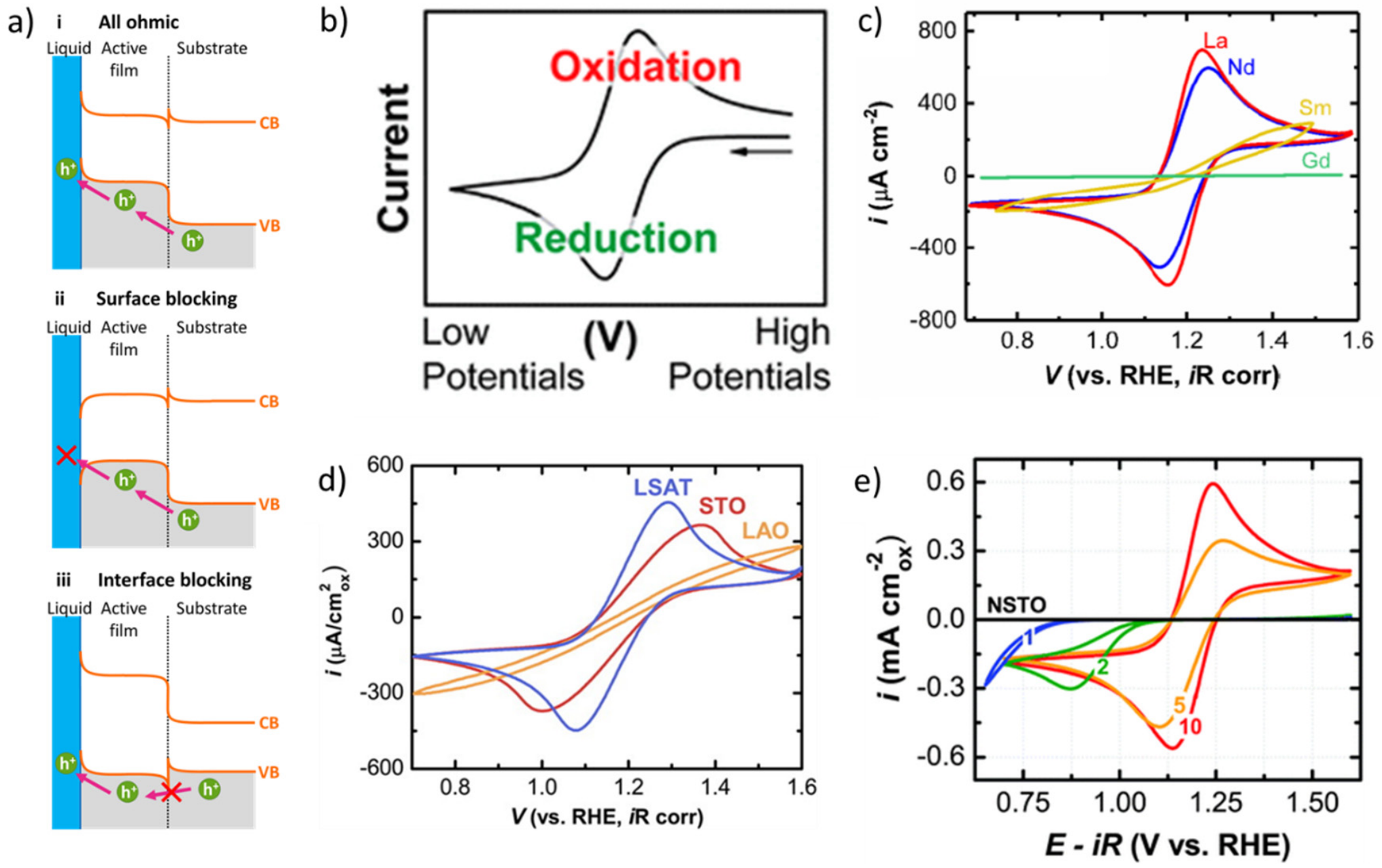

**Fig. 5.** Influence of the band structure on electrochemical charge transfer. (a) Schematic band structures showing (i) a non-blocking contact for hole transfer, (ii) transfer blocked at the liquid-solid interface and (iii) transfer blocked at an internal solid-solid interface. (b) Textbook example of a fast redox couple recorded on an electrode with metallic conduction. The corresponding measurements on epitaxial perovskite oxides may differ due to blocking interfaces for unsuitable (c) composition, (d) substrate or (e) film thickness. Panels were b, d were reproduced with permission from refs. [55,101]. Copyright American Chemical Society 2015, 2018. Panel c republished with permission of John Wiley and Sons from ref. [58]. Panel e republished with permission of the Royal Society of Chemistry from ref. [102].

The perovskite oxides reviewed herein are semiconductors and not metals, which is an important distinction for the necessary charge transfer from the solid to hydroxide/water in solution [103]. Yet, some perovskite oxides, e.g. ABO3 (ref), behave closely to p-type metals (refs). On metals, charge transfer is possible within a bandwidth of a few $k_BT$ (~25 meV near room temperature) from the Fermi level, i.e. it is thermally activated. In contrast, the Fermi level in intrinsic semiconductors falls between the band gap and charge transfer can only occur from levels within the valence or conduction bands, which requires higher energy. Furthermore, the charge carrier concentration in metals is usually not a limiting factor, while it can be a severe limitation for semiconductors. For the latter, the distinction between the majority carrier and the minority carrier (i.e., electrons or holes) is important. An n-doped semiconductor is one with the Fermi level (or localized states such as surface states) near the conduction band where electrons are the majority carriers (e.g., $Nb^{5+}$ ions on $Ti^{4+}$ sites in Nb:$SrTiO_3$), while a p-doped semiconductor has the Fermi level near the valence band and holes are the majority carriers (e.g. $La_{0.88}Sr_{0.12}FeO_3$ in Table 1). The concentration of the majority carriers may be similarly high to metals while the concentration of minority carriers certainly limits the charge transfer rate. Since the desired reaction here is the oxidation of hydroxide/water to oxygen, the relevant charge transfer is that of a hole from the surface of the solid to adsorbed hydroxide/water and high hole concentration is desirable.

The OER requires the transfer of four electron holes from the perovskite surface to hydroxide (alkaline) or water (neutral/acid). The composite electrodes discussed here have a liquid-solid interface between the electrolyte and the surface of the active perovskite as well as at least one buried solid-solid interface between the active surface and either a buffer layer or the substrate (Fig. 5a). With the exception of $La_{0.88}Sr_{0.12}FeO_3$, all oxides in Table 1 have bulk conductivities of at least 0.8 S/cm and can thus be considered good conductors. Yet, charge transfer of some of composite electrodes is hindered at one of the interfaces, which is independent of bulk conductivity and instead depends on the band gaps, the alignment of the bands and the relative electron affinities on either side of the interface. Since electron holes are transferred, we focus on the valence band (VB). Three possible cases are shown in Fig. 5a: (i) all Ohmic contacts where the holes can transfer across the internal interface and the surface; (ii) an Ohmic contact at the internal interface but a blocking Schottky contact at the surface that prevents hole transfer due to unfavorable band bending; and (iii) an Ohmic contact at the surface but a Schottky contact at the internal interface. Whether charge transfer across an interface or surface limits catalysis can be investigated using a fast redox couple with a single charge transfer as a charge acceptor simpler than the multi-step catalytic reactions such as the OER.

A convenient choice for investigations of charge transfer in alkaline electrolytes is the ferri-/ferrocyanide (FCN) redox with reversible potential of $E^0 \approx 1.2$ V vs. RHE that is similar to the reversible potential of water oxidation of $E^0 = 1.23$ V vs. RHE in alkaline media. (Note: $E^0$ of FCN depends on pH on the RHE scale while that of the OER does not). The textbook example of the CV trace of a fast redox could on a metal electrode is shown in Fig. 5b. It is symmetric about the reversible potential between the oxidation and reduction peaks. If the CV on a semiconductor looks comparable to that in Fig. 5b, then it behaves like a metal electrode that can both oxidize and reduce FCN, which cannot be assumed a priori for a semiconducting oxide. Additional complications may arise in epitaxial oxide films due to blocking behavior as discussed above. The two peaks in the textbook example have approximately identical currents above the baseline before the onset of the catalytical rise and the expected separation is 59 mV at 25 °C independent of slow sweep speeds (10's of mV/s) [104,105]. However, the latter value is rarely observed in aqueous solutions even when using metal electrodes [106–108], possibly due to interactions between the surface and FCN [109–111]. The reported values on perovskite oxides are in the range of 90 to 115 mV for the reversible CVs [59,102,112] and much larger if charge transfer across an interface is limiting.

The shape of the CV of FCN depends not only on the chemistry of the surface film, but also on its thickness and the choice of the substrate, where the latter two are specific to epitaxial oxide layers. In exemplary measurements on $ANiO_3$ (Fig. 5c), the peak currents of the FCN redox reduced in the order A = La, Nd, Sm, Gd. For Sm at the A-site, the shape is severely distorted with no discernable peaks and no currents were detected for Gd at the A-site [58]. The authors show that $SmNiO_3$ and $GdNiO_3$ had the lowest bulk conductivities in the series. Yet, additional interfacial barriers may exist, which would be desirable to quantify.

The effect of the substrate composition on the FCN redox was investigated by Stoerzinger et al. (Fig. 5d) [55]. An example is $LaCoO_3$ where the peak heights and separations increase from LSAT ($(LaAlO_3)_{0.3}(Sr_2AlTaO_6)_{0.7}$) to STO ($SrTiO_3$) to LAO ($LaAlO_3$). The curves remain symmetric, which indicates a reduction of available charge carriers rather than blocking due to bending bands at the surface. The authors argue that intermediate and high spin Co is formed due to tensile strain as reported previously [113]. It is plausible that this affects the number of available charge carriers, which has not been investigated systematically.

Stoerzinger et al. [102] also investigated the effect of $LaMnO_3$ film thickness on Nb-doped $SrTiO_3$ (Fig. 5c), where a 10 nm film showed the desired symmetric CV, while 5 nm films became slightly asymmetric with reduced currents and films below 2 nm could not oxidize FCN and showed severely lowered onset and currents for reduction. The authors attribute the deactivation to a reduction of $Mn^{3+}$ (in the bulk of $LaMnO_3$) to $Mn^{2+}$ on the surface and band bending on the surface. Due to the bulk Fermi levels, an electron-rich space charge layer at the surface (i.e. a reduced surface) was expected and the authors argue that only the space charge layer of the inactive Nb-doped $SrTiO_3$ formed for the thinnest $LaMnO_3$ films. While this sets a physical boundary to the minimal film thickness, the effect can be exploited to deposit very thin functional surface layers on active films, e.g., as demonstrated by Eom et al. [114] and Akbashev et al. [99].

In summary, we have discussed charge transfer from semiconducting electrodes to species in solution where blocking contacts or insufficient charge carriers may limit the charge transfer. This could mistakenly be attributed to the thermodynamics or kinetics of catalysis and it is therefore an important aspect to investigate before the interpretation of catalytic measurements. The experiments and their interpretation are straightforward and we urge everyone in the field to report this information for any new combination of perovskite film and substrate composition as well as the film thickness.

## 6. Activity trends

Activity metrics

Table 1 comprehensively lists all investigations of the OER using epitaxial oxides with perovskite or double perovskite structures, for which the electron transfer across the surface was not limiting as either probed by FCN redox or evidenced by significant OER currents. The perovskite structure was experimentally verified for all entries in the table. Common performance metrics of the OER are the currents normalized by the active area at a fixed voltage or the voltage, often overpotential, at fixed current density. All investigated films are very smooth so that the area exposed to the electrolyte can be taken as the oxide area. We evaluated the voltage at 50 $\mu A/cm^2_{ox}$ in our discussion, which has also been used previously [27]. A lower voltage (i.e. overpotential) is desirable for a good catalyst. Moreover, Table 1 lists the Tafel slopes of the perovskite oxides. The value of the Tafel slope depends on the electrons transferred before and during the rate-limiting step and also on the coverage of intermediates.[115,116] A full mechanistic discussion of the listed Tafel slopes is beyond the scope of this review, yet the Tafel slope presents the scaling of the current with the applied voltage and is thus also a performance metric. Their values range from 36 mV/dec ($NdNiO_3$ on STO) to 141 mV/dec ($LaNiO_3$ on LAO). Therefore, the catalytic trends that use pairs of voltage and current density are not universal and may dependent on the choice of the reference current density or voltage.

Investigations on substrates and hence films with (100) orientation are reported most frequently. Yet, there is a clear effect of the orientation of low index surfaces on the overpotential of the OER for $SrRuO_3$ films (Table 1) [66], $LaCoO_3$ films [71] and $La_{0.8}Sr_{0.2}CoO_3$ films [72] as well as single crystals [73]. For $SrRuO_3$, the reported overpotentials with orientation increase in the order (111) $\gtrsim$ (110) > (100) but the measurements suffered from Sr and Ru dissolution where the most active film dissolved the most cations. Nonetheless, the authors estimate about 90 % Faradic efficiency using RRDE measurements. Interestingly, this trend is not confirmed on related investigations of epitaxial rutile $RuO_2$ films where the reported current densities decrease with orientation in the order (100) > (101) > (110) > (111) [117,118]. It should also be noted that the trend of Ru dissolution differs, namely (111) > (101) > (100) > (110) [119].

The orientation dependence of perovskite oxides with Co at the B-site has also been studied. The current density of $LaCoO_3$ decreased with pseudocubic orientation in the order (100) > (110) > (111) [71]. In contrast, Sr doping changes the trend on $La_{0.8}Sr_{0.2}CoO_3$ to (110) > (100) ≈ (111) on both epitaxial films and single crystals [72]. Overall, there is no universal trend of activity with orientation of the epitaxial perovskite oxide. Based on the limited available data, both the A-site and B-site composition may influence the activity trends with orientation. Further investigations are desirable as it cannot be assumed that the commonly used (100) orientation is the most active one.

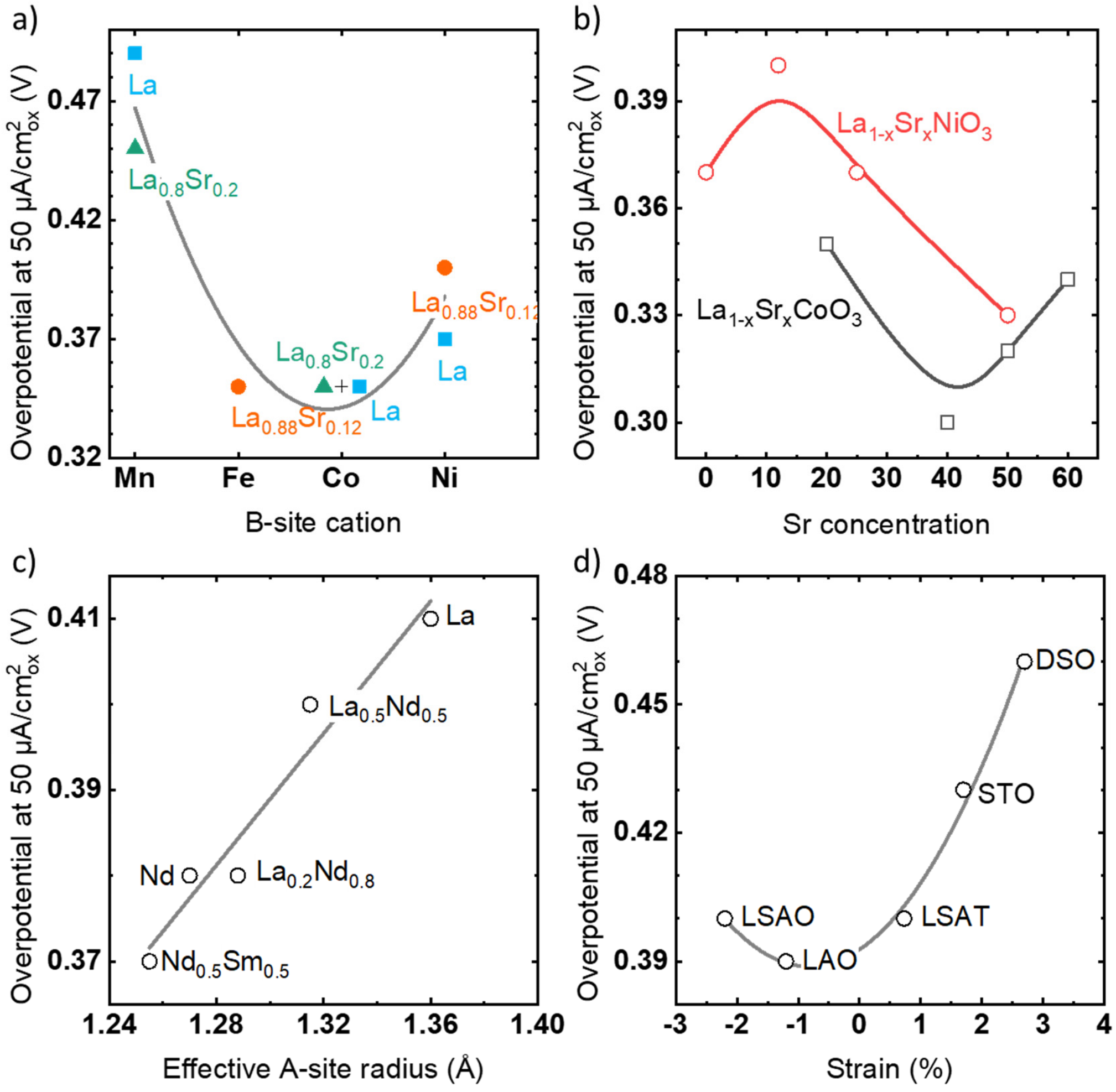

**Fig. 6.** Trends of the overpotential of the OER evaluated at 50 µA/cm$^2_{ox}$ with (a) the B-site cation, (b) the effective ionic radius of the A-site, (c) Sr-doping of $La_{1-x}Sr_xCoO_3$ and $La_{1-x}Sr_xNiO_3$, and (d) tensile or compressive strain of $LaNiO_3$ on different substrates. Lines serve as guide to the eye only. Note the different y-axis scales. The data were taken from refs. [28,55–58,61,62,65].

The reported B-site compositions on (100)-oriented films mainly focus on transition metals of the first row, namely, Mn (9 % in Table 1), Fe (8 % in Table 1), Co (32 % in Table 1) and Ni (45 % in Table 1). No systematic series has been reported with identical A-site and thus we included several A-site compositions in the comparison (Fig. 6a). The lowest voltage was found for Co- and Fe-based perovskites at fixed A-site composition (A = La, $La_{0.8}Sr_{0.2}$ or $La_{0.88}Sr_{0.12}$), which also includes the overall lowest overpotential of 0.24 V vs. RHE on $La_{0.8}Sr_{0.2}CoO_3$. Perovskite oxides of Ir and Ru have also been reported, where (111)-oriented $SrRuO_3$ may have an even lower overpotential of 0.07 V vs. RHE (Table 1) [66]. However, this comes at the cost of severe corrosion, which deactivates the film during the first cycle and may contribute to the current. The least active B-sites contain Mn. Earlier transition metals than Mn and later transition metals than Ni have not been reported yet extrapolating the reported trends suggests that they are also less active than B-sites of Co, Fe and Ni, which is also found in powder studies, e.g., in ref [27].

The majority of the studied A-site compositions contain La (26 % in Table 1) or La/Sr mixtures (25 % in Table 1). For the mixtures, the concentration of $Sr^{2+}$ at the A-site is often attributed to an increase of the concentration of $M^{4+}$ cations at the B-site and thus the B-site valence. The trend with the composition of the A-site was studied by Liu et al. [56] for $La_{1-x}Sr_xNiO_3$ ($0.0 < x < 0.5$) and by Stoerzinger et al. [57] for $La_{1-x}Sr_xCoO_3$ ($0.2 < x < 0.6$) as summarized in Fig. 6b. In both studies, the lowest overpotential was found between 40 and 50 % of the divalent Sr on the A-site, for which an A-site valence of +3.4 and +3.5 is expected. However, it should be noted that $La_{0.5}Sr_{0.5}NiO_3$ was the most Sr-rich composition investigated in the series with Ni at the A-site and that the activity per oxide area increased further with the maximum at $SrNiO_3$ with NiO and $SrCO_3$ impurities in a study based on powders [120]. La-rich compositions of $La_{1-x}Sr_xNiO_3$ are difficult to synthesize in pure phases [36,121–123], which may cause the discrepancies. Similar discrepancies arise when the trend reported for $La_{1-x}Sr_xCoO_3$ thin films is compared to the corresponding powders in a wider range, where the maximum is also outside the investigated parameter space at x = 0.8 [124] or x = 1.0 [125] and no local maximum of activity at x = 0.4 was found. In the epitaxial film study [57], this maximum was commensurate with the highest carrier concentration and mobility, which may differ between thin films and particles. Overall, these two studies suggest that changing the A-site composition results in more complex modifications than changing the ratio between $A^{3+}$ and $A^{4+}$ cations, which is often used to optimize the $e_g$ occupancy [124,126,127]. A comprehensive discussion of the impact of the A-site composition is beyond the scope of this review but in these two materials systems, undesired phase impurities may occur [120], the octahedral tilting and thus space group changes [125], the formation of oxygen vacancies may play a role [125,128], the carrier concentration as well as mobility may depend on doping [57] and the effective size of the A-site changes, which will modify the tolerance factor.

The influence of the size of the film's A-site cation on the catalytic activity was studied by Wang et al. for the perovskite oxides with $ANiO_3$ deposited on Nb:$SrTiO_3$ (Fig. 6c). The highest overpotential of 0.41 V at 50 µA/cm$^2$ was observed for $LaNiO_3$. The overpotential decreases when La is systematically replaced by Nd, where the lowest overpotential was found for $NdNiO_3$. An additional reduction of the overpotential to 0.37 V was achieved by replacing 50 % of Nd with Sm. The authors attributed the observed trend to the reduction of $Ni^{3+}$ ($e_g^1$) to $Ni^{2+}$ ($e_g^2$) and the concomitant formation of oxygen vacancies. These substitutions also decrease the tolerance factor from La ($t$ = 0.996) to Nd ($t$ = 0.958) and thus make the more active perovskites less cubic. This puts a natural boundary to increasing the activity with this approach and the activities in the study showed negligible effects when the effective A-site radius is below 1.28 Å, which suggests that the trend is not continued by further decreasing the effective A-site.

Strain between the substrate and epitaxial films affects many physical properties [119,129–132] and likewise the catalytic activity as investigated for various combinations of substrates and perovskite oxides based on Mn [100], Co [55,74] and Ni [54,61] (Fig. 6d). In the shown example for $LaNiO_3$, the lowest overpotential is found for the least strain where the overpotential increases likely quadratically with positive and negative strain. Thus, minimal strain is strongly recommended for $LaNiO_3$ films. Yet, for $NdNiO_3$ films, the activity increases with either compressive or tensile strain [54]. The authors argue that the former is beneficial for OER due to an enhanced p-d hybridization [133] and the latter due to the formation of oxygen vacancies and the concomitant reduction of Ni, which increases the average $e_g$ occupancy to slightly above unity, i.e. closer to the optimal value [27]. The effect of strained films has also been studied for Co at the B-site. Stoerzinger et al. [55] found that slightly tensile strain (+1.8%) resulted in the fastest kinetics for charge transfer (Fig. 5d) and also the highest OER activity. The authors argue that tensile strain increases the Co-O bond strength at fixed $e_g$ occupancy. In contrast, Petrie et al. [61] found that the highest tensile strain of +4.2% resulted in the highest OER activity due to the formation of oxygen vacancies. In summary, the effect of strain depends on the active film and changes multiple materials properties that had been previously correlated with OER activity. Therefore, the effect of strain is difficult to predict and the compositional (e.g. oxygen vacancies) or electronic changes (e.g. hybridization) must be considered to identify the dominant effect on the OER.

The complex effect of seemingly simple substitutions in epitaxial perovskite oxide films on perovskite substrates has been discussed from various angles in this review. As stated above, we here focus on epitaxial systems which are not (strongly) limited by charge transfer. While perovskites are good materials system for systematic investigations, the caveat is that compositional changes affect multiple properties, which may all impact the OER. This is in line with the statistical evaluation of Hong et al. [33] who likewise found that multiple properties must be considered for strong predictive relationships, where electron occupancy and metal-oxygen covalency are the dominant factors but structural influences such as M-O-M angles (i.e. octahedral tilt) and the tolerance factor also matter. In our examples in Fig. 6, the strongest effect on the OER was observed with B-site substitution (110 mV; Fig. 6a), then A-site composition and strain (both 70 mV; Fig. 6b,d), and finally substitution of the A-site (40 mV; Fig. 6c). These trends agree with the ranking performed for systematic studies of perovskite oxides catalyzing the oxygen reduction reaction (ORR), where electronic changes on the B-site also had the largest impact and changes on the A-site had the least impact [93]. Yet, these properties are not general enough to be easily transferred to other perovskites or other materials classes, which motivates the search for predictive descriptors of activity that relate to mechanistic details, most commonly oxygen adsorption, which is a mandatory step for the OER.

Rational design of electrocatalysts for the OER requires fundamental insight into the microscopic origin of activity, which can be derived from structure–activity relationships. In oxygen electrocatalysis, this is frequently called the descriptor approach [23,27,36,28–35], where some physical property of the material serves as the descriptor for catalytic activity. Many experimental descriptors have been proposed and tested for the oxygen evolution reaction. Popular choices include electronic properties such as the number of d electrons [36], the number of outer electrons [30], the valence state, the number of $e_g$ electrons [27], the O-p band [97] and the metal-oxygen covalency [94,134]. Similar to descriptors for the ORR [93], reports of structural descriptors [33,135] and magnetic descriptors [33,136] are very scarce. All of the experimental structure-activity relationships employ powders rather than epitaxial films, which naturally leads to the question: How well do popular descriptors of the OER work for epitaxial perovskite systems?

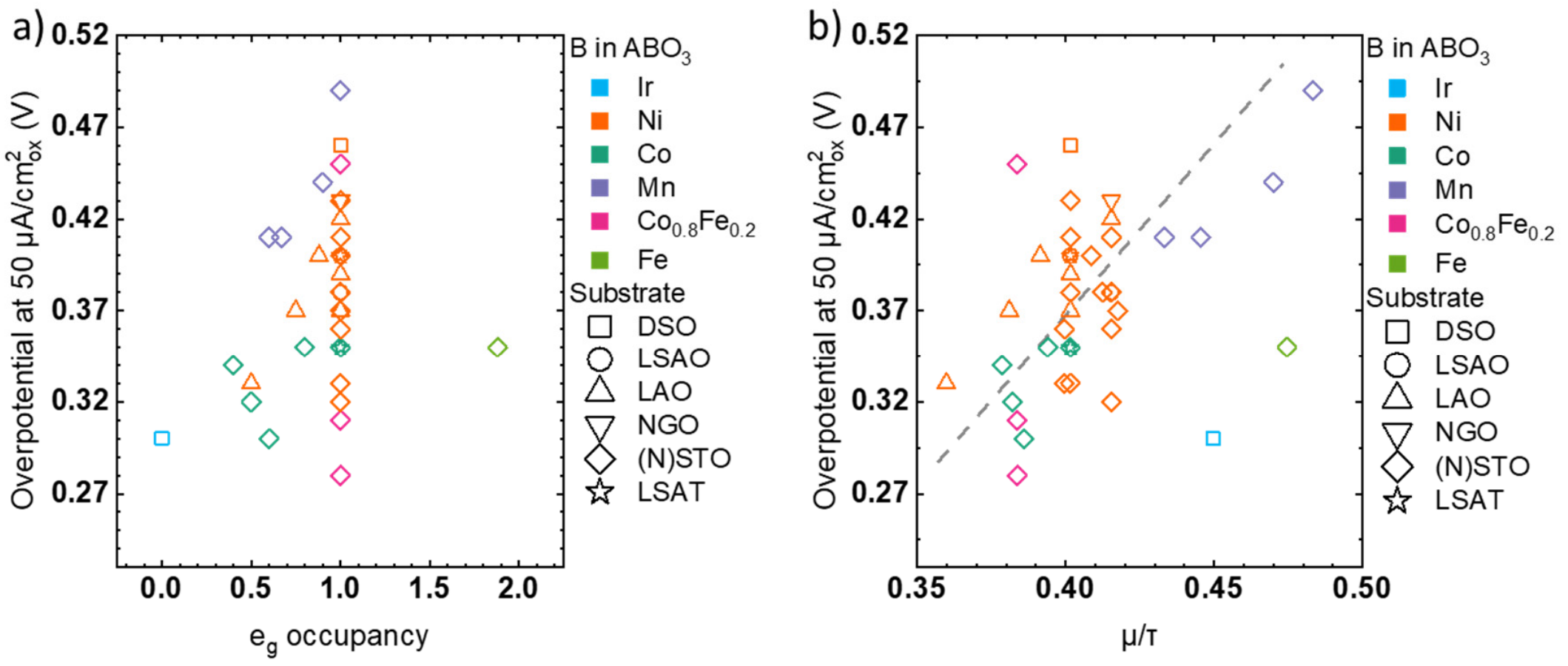

**Fig. 7.** Overpotential at 50 µA/cm$^2$ of perovskite oxides in 0.1 M hydroxides as function of (a) a chemical descriptor ($e_g$ occupancy) and (b) a structural descriptor ($\mu/t$). The dashed line is a subjective guide to the eye. The symbol shape indicates the substrate and the color the B-site metal. The data was taken from refs. [28,53,62–70,54–61].

We selected the $e_g$ orbital occupancy and the ratio of the octahedral to the tolerance factor (µ/t) as an electronic and a structural descriptor to plot the data in Table 1 (Fig. 7). The analysis was limited to 0.1 M hydroxides to exclude pH effects that have been reported for perovskite oxides [128,137,138]. The $e_g$ occupancy was proposed by Suntivich et al. for the ORR [95] and then also the OER [27] on perovskites oxides. It has since gained much popularity and also applies beyond the perovskite structure, e.g. for spinels [139]. The $e_g$ orbital points toward (end-on) adsorbed oxygen and thus correlates with oxygen adsorption as an elementary step in the mechanism of the OER. The ratio of the octahedral to the tolerance factor was recently proposed by Weng et al. [135] based on symbolic regression. As these quantities are only defined for perovskites and related materials, the descriptor is less general than the $e_g$ occupancy and it does not clearly relate to oxygen adsorption or another elementary mechanistic step. We believe that it most likely relates to tilting of the octahedra and as such to the metal-oxygen covalency and charge transfer. Nonetheless, its value can be easily calculated from readily available tables [51] and Weng et al. [135] showed that it scales linearly with the OER activity.

We could not identify any apparent trend of the reported overpotentials at 50 µA/cm$^2$ data with $e_g$ occupancy (Fig. 7a). The $e_g$ occupancy was calculated based on the reported stoichiometries where the most common spin state (high spin, hs, intermediate spin, is, or low spin, ls) was assumed. i.e. hs $Mn^{3+}$, hs $Fe^{3+}$, hs $Fe^{4+}$, is $Co^{3+}$, hs $Co^{4+}$, ls $Ni^{3+}$, ls $Ni^{4+}$. In the shown representation, a V-shaped curve (inverted volcano) with the minimum near unity $e_g$ occupancy would be expected based on the previous works on perovskite oxide powders. However, no clear trend with $e_g$ occupancy could be identified. Naturally, a large spread should be expected because the data were recorded by different groups with different protocols, because the films were made by different methods and most importantly because the aim of the investigations was often to demonstrate the effect of other parameters at nominally fixed $e_g$ occupancy, e.g., strain (different symbols in Fig. 7a). Our plot should not be misunderstood that the $e_g$ occupancy is unsuitable as a descriptor, which was proven for powders [27]. Instead it illustrates the limits of the $e_g$ descriptor. Many of the reports included in our plot discussed qualitative deviations from the formal valence [53,54,68,55–59,61,65,67], In that sense, the $e_g$ descriptor is not very robust against guessing it from the formula and it should be quantified consistently by measurement for films made by identical deposition methods. As a final remark, the spread of the data at $e_g$ occupancy of one (210 mV) as compared to the span of overpotentials in Fig. 7a (190 mV) suggests that descriptors other than the $e_g$ occupancy can have an equally large impact on the OER activity of epitaxial perovskite oxides, which is in contrast to powders studies [27,93,95].

We also plotted the same overpotentials at 50 $\mu A/cm^2_{ox}$ as function of $\mu/t$ (Fig. 7b). The tolerance ($t$) and octahedral ($\mu$) factors were calculated using Eq. (1) and (2) with the radii obtained from Shannon's table [51]. Twelve-fold coordination was assumed for all A-sites. Surprisingly, the linear trend expected from the work of Weng et al. [135] could also be found for epitaxial perovskite oxides on various substrates, albeit with large scattering. Overall, it can be concluded from our plot that a smaller $\mu/t$ most likely leads to lower overpotentials for (mainly) first row transition metals on the investigated substrates. Notable outliers are $La_{0.88}Sr_{0.12}FeO_3$ with low bulk conductivity and $SrIrO_3$. Upon closer inspection, the spread at selected values of $\mu/t$ is still considerable, e.g., 140 mV at $\mu/t$ = 0.40. We expect that it has the similar origins as discussed above for the $e_g$ descriptor, which result in modifications of the bond lengths due to strain and due to octahedral tilting (Fig. 1a,d). In contrast to the $e_g$ descriptor, the magnitude of the secondary effects (estimated as the spread of 140 mV at $\mu/t$ = 0.40) is smaller as compared to the span of overpotentials in the dataset (210 mV). Experimental determination of the of the actual bond lengths would be desirable to obtain a more accurate $\mu/t$ descriptor. Overall, we found that the structural μ/t descriptor was more robust against guessing its value from the formula and tabulated values as compared to the popular electronic descriptor of $e_g$ occupancy for epitaxial perovskite systems.

A final reason for deviations from trends may be chemical or structural changes during the OER, which is still rarely reported. Stability of the perovskites at voltages below the onset of OER and during the OER is a topic to extensive to be covered herein. Instead, we give a brief overview. It has not been rigorously established that there is a perovskite surface that does not change during the OER. The reported modifications range from change in the surface chemistry with retained perovskite framework to surface reconstruction to complete loss of crystalline structure [59,100].

Scholz et al. [59] reported that the surface of $La_{0.6}Sr_{0.4}MnO_3$ is enriched in La on the surface due to Mn and Sr dissolution. The surface remained in the perovskite structure as witnessed by TEM. These films showed negligible changes in XRR, XRD and AFM. Weber et al. [64] also found a similar La enrichment on the surface of $La_{0.6}Sr_{0.4}CoO_3$ after electrocatalysis due to B-site (Co) and Sr leaching. Their data indicates a loss of the perovskite structure to an amorphous structure as also reported for Co-based perovskite oxide powders [140,141]. This transformation shows clear changes in both XRD and AFM. The induced changes depend on the number of cycles, i.e. the passed charge [62,64], and the voltage range of the cycles [69].

Environmental TEM (ETEM) studies can give direct insight into surface changes but can currently only be performed in water vapor. Mildner et al. [142], Mierwaldt et al. [143] and Roddatis et al. [67] discuss the surface reconstructions of $Pr_{1-x}Ca_xMnO_3$ lamella when exposed to water vapor. Interestingly, the amorphization sometimes observed for on the surface of perovskite oxides can be reversible, e.g. preparation-induced amorphous regions on the surface recrystallized in the presence of water on $Pr_{0.64}Ca_{0.36}MnO_3$ [67,142,143]. In contrast, $Pr_{0.9}Ca_{0.1}MnO_3$ remained crystalline in water vapor but showed cation dynamics on the surface [143].

The most extreme changes have been reported by Chang et al. [66] for some of the most active perovskite in Table 1, namely $SrRuO_3$. Similar to $La_{0.6}Sr_{0.4}MnO_3$ and $La_{0.6}Sr_{0.4}CoO_3$, the B-site (Ru) and Sr are dissolved, while for the Ru-based perovskites drastic changes were evident from the cyclic voltammogram during the first cycle as the current becomes erratic soon after the exponential rise due to OER. Additionally, post mortem analysis of the Bragg peaks (crystal truncation rods) corresponding to the surface shows dramatic changes that indicate severe roughening (i.e. extreme surface reconstruction).

In summary, leaching of the B-site transition metal and group II A-site cations appears to be common but we caution that very few perovskites have been investigated. Enrichment of the surface with La was reported for those perovskites with La and Sr on the A-site. The degree of surface reorientation varies between dynamic cation movement but no changes in a time average to complete irreversible loss of the perovskite structure, to which the more active perovskites are prone.

## 9. Concluding remarks and outlook

In this review, we have summarized the current state-of-the art of investigations on perovskite oxides catalyzing the OER. The perovskite structure is clearly defined and the understanding of the physics and chemistry of epitaxial oxides is very mature. Likewise, electrocatalytic investigations are well established, even though there is a dire need for a unified testing protocol for OER investigations to make the results more comparable. The challenge lies in consolidating the knowledge in these two fields and addressing the challenges that arise when epitaxial perovskite oxides are used as electrocatalysts for the OER. We introduced the prerequisites for charge transfer across interfaces, namely both a suitable alignment of the relevant energy levels and sufficient charge carriers. These parameters are not a concern for more classic electrochemical electrodes made of metals, but they are critical parameters for semiconductors such as perovskite oxides. These points are frequently discussed in the neighboring field of photoelectrocatalysis [144–149] but receive comparably little attention in the electrocatalysis of the OER. It would also be desirable to include the effects of energy alignments and charge carriers in theoretical calculations of adsorption energies used as catalytic descriptors (e.g. $\Delta G$ of OH). It is conceivable that the surface with optimal adsorption energy of a key intermediates such as OH, could have either unsuitable band edges for the OER or insufficient charge carriers. The need to align energy levels does not necessarily impact catalysis negatively. Eom et al. [114] and Akbashev et al. [99] used atomically controlled multi layers to optimize adsorption properties, stability and the alignment for optimal charge transfer. The prospects of these efforts have also been highlighted by Weber and Gunkel in this Special Issue [150].

In situ and operando spectroscopy and microscopy on epitaxial thin films is extremely challenging but desperately needed to address the key questions of the nature of the active state for the OER and its potential evolution during the OER. AP-XPS studies on epitaxial perovskite oxides have been key to understanding surface hydroxylation [57,151–153]. In situ X-ray diffraction has been used to study the structure and its evolution with potential for epitaxial oxides [154–156] but not perovskite oxides. In situ X-ray spectroscopy is used to elucidate the valence during OER for other catalysts but has not been used for epitaxial thin film oxides [157–160]. These in situ measurements are instrumental to understand the OER on epitaxial perovskite oxides and fully harness their potential as model electrodes to derive guidelines to improve their electrocatalysis by optimizing the composition or by exploiting new concepts inherent to epitaxial layers.

**Acknowledgments**

We thank Dr. Ibrahim Y. Ahmet for discussions. This project has received funding from the European Research Council (ERC) under the European Union's Horizon 2020 research and innovation programme under grant agreement No 804092.